\documentclass[pre,a4paper,oneside,twocolumn]{revtex4}

\usepackage[english]{babel}
\usepackage[latin1]{inputenc}
\usepackage{graphicx}  
\usepackage{latexsym}  
\usepackage{amssymb,amsmath,amsthm,amsfonts} 
\usepackage{color} 

\begin{document}

\title{A Non-Phenomenological Model to Explain Population Growth Behaviors}


\author{ \firstname{Fabiano}  \surname{Ribeiro}}
\email{fribeiro@dex.ufla.br}
\affiliation{Departamento de Ci\^encias Exatas (DEX), \\
                Universidade Federal de Lavras (UFLA) \\
	           Caixa Postal 3037 \\
		     37200-000 Lavras, Minas Gerais, Brazil}

\date{\today}

\begin{abstract}
This paper proposes a non-phenomenological model of population growth that is based on the interactions between the individuals that  compose the system. 
It is assumed that the individuals interact cooperatively and competitively. As a consequence of this interaction,  it is shown  that some well-known phenomenological population growth models (such as the Malthus, Verhulst, Gompertz, Richards, Von Foerster, and power-law growth models) are special cases of the model presented herein. 
Moreover,  other ecological behaviors can be seen  as the emergent behavior of such interactions. For instance,  the Allee effect, which is the characteristic of some populations to increase the population growth rate at a small population size, is observed. 
Whereas the models presented in the literature explain the Allee effect with phenomenological ideas, the model presented here explains  this effect by the interactions between the individuals.  
The model is tested with empirical data to justify its formulation.  
Other interesting macroscopic emergent behavior from the model proposed here is the observation of a regime of population divergence at a finite time. It is interesting that this characteristic is observed in humanity's  global population growth. It is shown that in a regime of cooperation, the model fits very well to the human population growth data since 1000 A.D.
\end{abstract}

\keywords{Complex Systems, pacs89.75.-k; 
Population dynamics (ecology), pacs 87.23.Cc;  
Pattern formation ecological, pacs 87.23.-n. }

\maketitle

\section{Introduction}\label{introduction}

The study of population growth is applicable to  many areas of knowledge, such as  biology, economics and sociology \cite{murray, solomon, polones, primate-societies}.
In recent years, this wide spectrum of applicability has motivated 
a quest for universal growth patterns that could account for different  types of systems by means of the same idea  \cite{chester, guiot, west-nature, west-pnas}.
To model a more embracing context, generalized growth models have been  proposed to address different systems without specifying functional forms \cite{generalized_model_gross}.
These generalized models have  helped guide the search for such  universal growth patterns \cite{fabiano-pre, fabiano-physicaA, fabiano_2species, vector_growth}.

The first population growth models were proposed to describe a very simple context or a specific empirical situation.  
For instance, the \textit{Malthus model} \cite{malthus, murray} was proposed to explain populations whose growth is strictly dependent on the number of individuals in the population, i.e. populations that have a constant growth rate. The model yields to an exponential growth of the population, and although it fits very well to some empirical data when the population is sufficiently small, it fails after a long  period of time \cite{livro_math_bio, murray}.  
To describe a more realistic  population, 
Verhulst introduced \cite{verhulst-1845, verhulst-1947, murray} a quadratic term in the Malthus equation to represent an environment with limited resources.
The \textit{Verhulst model} yields to the logistic growth curve, which fits  many empirical data very well; examples includes bacterial growth and human population growth \cite{livro_math_bio, murray}. 
Another important model is the \textit{Gompertz model}, which was introduced in \cite{gompertz} to describe the human life span but has many others applications \cite{gompertz_1998}. The model is a corruption of Malthus's original model by the substitution of a  constant growth rate with an exponentially  decaying growth rate \cite{ausloos}. The model yields to a asymmetric sigmoid growth curve.

In the last few decades,  a search for theoretical models that aggregate as many situations as possible has been conducted; the idea is that the larger the applicability of the model is, the better the theory is \cite{chester}. 
For instance, the \textit{Richards model}, which was introduced to describe plants' growth dynamics \cite{richards_59, richards_1998},  has the Verhulst and Gompertz growth models as particular cases. 
Another  model important in this context is the \textit{Bertalanffy model} \cite{bertalanffy-paper, bertalanffy-model, savageau-1979}, which summarizes  many classes of animal growth using the same approach. An additional model, which was introduced in \cite{polones_q}, presents a generalization of the Malthus and Verhust models based on the generalized logarithm and exponential function  \footnote{The generalized forms of the logarithm and exponential function are discussed in the appendix \ref{appendix_lnq}}.
Other types of models that deserve attention are the ones that use an expansion of the Verhulst therm in a power series and apply it  to multiple-species systems \cite{generalized_model_gross, solomon}. 
Furthermore, there are models that use second-order differential equations to describe growth, and these models have been strongly corroborated by empirical data \cite{chester,second-order-EDO}.

All of the models cited above can be seen as \textit{phenomenological models},  because the only assumption that such models take into account in their formulation is the population's - macroscopic level - information. This information includes, for example, the population's  size, density, and  average quantities. The particularities of the individuals - the microscopic level -  are removed from the formulation of these models. 
It is the thought process of most of the models presented in the literature. 
This approach is very appropriate, as it is difficult to know in detail the particularities of all of the components of the population.  Indeed, taking these details into account complicates the calculus and computations that are  necessary to predict  the population behavior from the model.
However, finding universal patterns of growth is extremely helpful in observing  how the components of the system behave. 
There would most likely be some types of individual behaviors that  are common even in different systems. If that is the case, then  one can justify the same pathern of growth being  observed in completely different type of systems as a consequence of similarities at the microscopic level. 

It is observed  in many fields of science that  simple interaction rules of the components of a system can result  in complex macroscopic behavior.
Moreover,  some properties of such systems are universal, such as the same \textit{critical exponents} in magnetic and fluid systems; 
these properties are universal even in systems that are completely different \cite{yeomans, kodanoff}. 
In the language of \textit{complex system theory}, it is said  that the collective  behavior (macroscopic level) \textit{emerges}  from the interactions of the components of the system (microscopic level). 
Thus, the collective effects are  called  \textit{emergent behavior} \cite{boccara, mitchell}. 
The idea of Mombach et all, that is reported in \cite{mombach}, which will hereafter be referred to as the \textit{MLBI model},  was to apply  emergent behavior's idea  to population growth. 
Hence, in oppositione to the common models that present modeling from a phenomenological point of view, this model is based on microscopic assumptions. 
As a result, the (non-phenomenological) MLBI model, which was formulated in the context of inhibition patterns in cell populations,  
reaches many well known phenomenological growth models (such as the Malthus, Verhulst, Gompertz and Richards models) as an emergent behavior from individuals' interactions. Recently,  this model was analyzed in \cite{donofrio, fabiano_2species, fabiano-physicaA, fabiano-pre}.

The model that is proposed here continues the main idea of the MLBI model. 
However, the proposal of the present work is to increase the scope of this model. It will be considered that the individuals that  constitute the population interact with each other not only through  competition, as was proposed in the original MLBI model, but also through  cooperation.
The emergent behavior of this more embracing formulation is the \textit{Allee effect}, which is the property   of some biological populations to increase their  growth rate with increases in the population size for small population. 
This behavior cannot be  deduced from the original MLBI model.
Other interesting macroscopic emergent behavior from the model proposed here includes  the observation of a regime of population divergence after a finite amount of time.  It is interesting that this  characteristic is observed in  humanity's global population growth, as will be shown in the following sections of the paper.  

The paper is organized as follows:  
In the first section, a model based on the interactions (cooperation or competition) between the individuals of the population is presented,  and in the second section, it will be shown that the model can explain the Allee effect in a non-phenomenological way. That is, the Allee effect can be explained by the interactions between the individuals of the population. The model is tested with empirical data to justify its formulation. 
In the third section, it  will be shown that some very important models in the literature can be obtained by changing some  variables of the present model, such as the strength of the interaction, the geometry in which the population is  embedded,  and the spatial distribution of the population. Thus, some very-know models in the literatur can be seen as special cases of the present model.

\section{The Model}\label{section_model}

The work presented here is based on the MLBI model, which was introduced in  \cite{mombach} and reworked by D'Onofrio in \cite{donofrio}.
The MLBI model was proposed to explain the population growth of cells by considering the inhibitory interactions between them. As a result,  researchers dicovered  that some very-known phenomenological models present in the literature (such as Verhulst, Gompertz and Richard's models) can be obtained as consequence of the microscopic interactions between individuals.

The present work follows the idea of the MLBI model,  and expands its  applicability to other ecological systems. 
In particular an expanded version of the model is presented, and cooperative interactions between  individuals are introduced. 
Then, one shows that the model can deal not only with populations of  cells (the context that inspired the original version of the model), but also with the population growth of some mammals and even human populations. Moreover,  the expanded version of the model can explain the Allee effect, which is is not possible by regarding only  its original version.

First, consider that the replication rate $R$ of a single individual in a population is given by 
\begin{eqnarray}\label{eq_text}
R &=&  \text{\small{[Self-stimulated replication]}} - \text{\small{[competition from field]}}+\nonumber \\  & &  + \text{\small{[cooperation from field].}} 
\end{eqnarray}

\begin{figure}[htbp]
\centering
\includegraphics[width=\columnwidth]{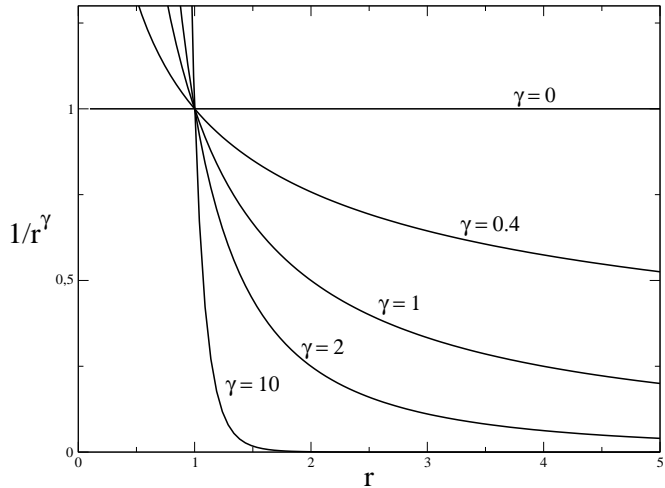} 
\caption{ \label{figure_decay} A graph that represents the intensity of the interactions between two individuals as a function of the distance $r$,  according to some values of $\gamma$. As  $\gamma$ (the exponent decay) increases,  the interaction range decreases. When $\gamma=0$,  the interaction range is infinity; that is, the intensity of the interaction does not depend on the distance.}
\label{graph-int-r}
\end{figure}

This idea is identical to the one proposed by Mombach (in the first line of (\ref{eq_text})),   except for the cooperative stimulus term (the second line).
Following Mombach et al, suppose that the intensity of the competitive interaction between two individuals decays according to  the distance $r$ between them in the form $1/r^\gamma$, where  $\gamma$ is the \textit{decay exponent} (see figure~(\ref{figure_decay})). In the present work, it will be assumed  that the cooperative interaction behaves in the same way. 
Thus, the replication rate of the $i$-th individual of the population has the form

\begin{equation}\label{equation_Ri}
 R_i = k_i - J_1 \sum_{j \ne i} \frac{1}{|\mathbf{r}_i - \mathbf{r}_j |^{\gamma_1}} + J_2 \sum_{j \ne i} \frac{1}{|\mathbf{r}_i - \mathbf{r}_j |^{\gamma_2}}. 
\end{equation}
This equation  mathematically represents the idea introduced in  equation~(\ref{eq_text}), where: 
$k_i$ is the self-stimulated replication  of the $i$-th individual; $\mathbf{r}_i$ and $\mathbf{r}_j$ represent the position vectors of the individuals $i$ and $j$, respectively, and consequently $|\mathbf{r}_i - \mathbf{r}_j |$ is the absolute distance between them;
$\gamma_l$ is the exponent decay of the competitive  interaction ($l=1$), or the cooperative interaction ($l=2$); 
lastly,  $J_l >0$ represents the strength of the competitive  ($l=1$) or cooperative($l=2$) interaction.

The update of the population must obey the rule  $N(t+~\Delta t) = N(t)+~\Delta t\sum_{i=1}^N R_i$. 
In the limit $\Delta t \to 0$, one has the differential equation $dN/dt = \sum_{i=1}^N R_i$. By equation~(\ref{equation_Ri}),  one has

\begin{equation}\label{equation_dNdt}
 \frac{d}{dt}N = \sum_{i=1}^N \left(k_i - J_1 I_i^{(1)} +J_2 I_i^{(2)}    \right),
\end{equation}
where 

\begin{equation}\label{defining_I}
 I_i^{(l)} \equiv \sum_{j\ne i} \frac{1}{|\mathbf{r}_i - \mathbf{r}_j|^{\gamma_l}} 
\end{equation}
is the (cooperative or competitive) interaction term that individual $i$ feels from its neighbors.

In appendix~(\ref{appendix_I}), a extended version of the calculus of Mombach et all with respect to the sum in equation~(\ref{defining_I}) is presented. One can prove (see the appendix) that the interaction factor $I_i^{(l)}$ is the same for all individuals of the population (regardless of $i$), and it has the form

\begin{equation}\label{equationI}
 I^{(l)} \equiv I_i^{(l)} = \frac{\omega_D}{D_f(1- \frac{\gamma_l}{D_f})}\left[ \left(\frac{D_f}{\omega_D}N \right)^{1- \frac{\gamma_l}{D_f}} -1 \right].
\end{equation}
Here, $\omega_D$ is a constant that depends exclusively on the  Euclidean dimension $D(=1,2,3)$ in which the population is embedded, and $D_f$ is the fractal dimension of the spatial structure formed by the population.
To maintain the physical propert that $I_i^{(l)}$ is positive (note that the sum in equation~(\ref{defining_I}) is over absolute terms), the model must be restricted to the case where  $\omega_D/D_f < N$. In fact, it is demonstrated  in the analysis around equation~(\ref{eq3}) (appendix~\ref{appendix_I}),  that  $\omega_D/D_f \sim 1$  (which is much smaller than the population size).

As presented in \cite{fabiano-pre}, one can write the term on the right-hand side of expression~(\ref{equationI}) by means of the generalized logarithm (see appendix~(\ref{appendix_lnq})):

\begin{equation}\label{equatin_I_lnq}
I^{(l)} = \frac{\omega_D}{D_f} \ln_{\tilde{q}_l} \left( \frac{D_f}{\omega_D} N \right),
\end{equation}
where  

\begin{equation}\label{definition_q}
\tilde{q}_l\equiv 1- \gamma_l/D_f.
\end{equation}
The parameter $q_l$ gives information about the relation between the decay exponent and the fractal dimension of the population. 

By introducing the average of the intrinsic growth rate $\langle k \rangle \equiv (1/N) \sum_{i=1}^N k_i$, employing the definition $J_l'\equiv J_l \omega_D/D_f$, and using result~(\ref{equatin_I_lnq}), one obtain from~(\ref{equation_dNdt}) the \textit{Richard-like} model 

\begin{equation}\label{equation_macro}
G(N) = \frac{1}{N} \frac{d}{dt}N = \langle k \rangle -  J_1'\ln_{\tilde{q_1}} \left( \frac{D_f}{\omega_D} N \right) +  J_2' \ln_{\tilde{q_2}} \left( \frac{D_f}{\omega_D} N \right).
\end{equation}

The \textit{per- capita growth rate}  $G(N)$ 
gives  information about the type of interaction that predominates in the   population. For instance,  $dG(N)/dN > 0$ means that cooperation predominates: the larger the population is,  the  larger the per- capita growth rate is. However,  $dG(N)/dN < 0$ means that competition predominates: the larger the population is, the smaller the per- capita growth rate is.

%

\subsection*{Comments}

The result~(\ref{equation_macro}) depends only on the macroscopic parameters of the system, besides to be deduced from  microscopic (or individual level) premises. 
That is a remarkable result, and it was first obtained by Mombach et all in the context of inhibitory interactions and is now expanded to cooperative interactions.
This result represents a significant  advance in the knowledge of patterns in population growth. 
It is because the model is not a phenomenological one, that is, it is not a model that is constructed to fit macroscopic data. The MLBI model, which was extended here,  is deduced from the own individuals' interactions. Then, the macroscopic behavior emerges as a consequence of the interactions.

Moreover, the model presented  in this work is more robust than 
its original version.  The original model is fully obtained by assuming that $J'_2=0$ in~(\ref{equation_macro}). In this case, the per- capita growth rate $G(N)$ 
is a monotonically decreasing function of the size of the population (because  if $J'_2=0$, then  $dG/dN <0$ for any population size).
Consequently, the simpler form of expression~(\ref{equation_macro}),   which does not  present the cooperative effects, cannot explain the \textit{Allee effect}. 
However, if the cooperative term is considered ($J'_2 \ne 0$ in~(\ref{equation_macro})),  the Allee effect can be predicted by the model. 
This effect will be discussed in more detail in the next section.

\section{The Allee Effect}

\begin{figure}[htbp]
\centering
\includegraphics[width=\columnwidth]{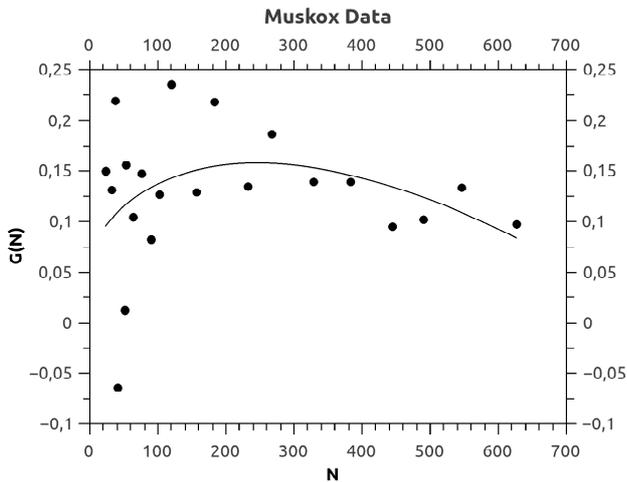} 
\caption{\label{fig_allee_data} The experimental data of the per- capita growth rate of a population of muskoxen  as a function of the population size. The data were obtained directly from \cite{gregory}.
The line is the fit from model~(\ref{equation_macro}) with the following parameters:
$J_1' \approx1.0305$;     
$J_2'\approx 1.0327$;     
$q_1 \approx 0.9859$; 
$q_2 \approx 0.9855$;   
and $\langle k \rangle \approx 6.828$.
The term $D_f/\omega \equiv 1$ was kept fixed. 
The statistical analysis shows $R^2= 0.10694$. }
\end{figure}

The Allee effect is the property of some populations to increase their  per- capita growth rates with  incriasing population size when the population is small \cite{allee_revisao, theta-logistic}. 
For instance, in figure~(\ref{fig_allee_data}), the  experimental data of the per- capita growth rate of the muskox population are presented  as a function of the population size.  
For a small population the experimental data show an increasing trend  with respect to the size of the population. Then, when the population is sufficiently large, the per- capita growth rate decreases.

This effect has been studied in recent theoretical models \cite{theta-logistic, courchamp1999, alexandre_allee}. 
However,  these models are restricted to the macroscopic approach and do not consider the microscopic level of the system.  
In the model proposed here, the Allee effect can not only be obtained,   but it can also  be interpreted as a macroscopic behavior that is observed as a consequence of the interactions of the individuals that  compose the population.

\begin{figure}[htbp]
\centering
\includegraphics[width=\columnwidth]{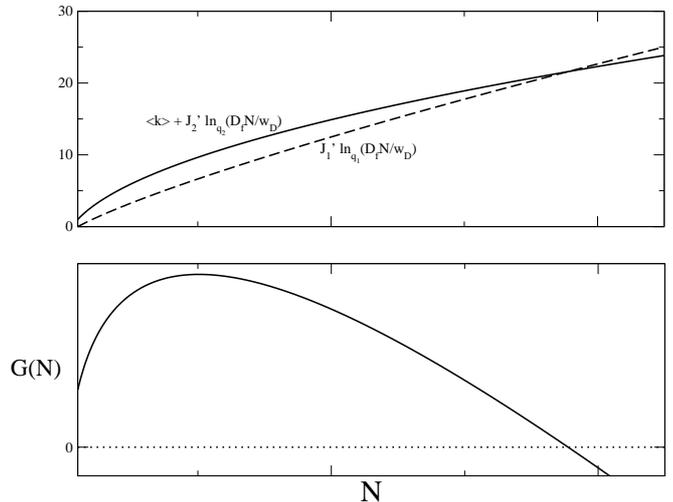} 
\caption{\label{fig_allee2} 
The upper graph presents separately the functions that compos $G(N)$ as a function of $N$, which conform to   
equation~(\ref{equation_macro}): 
$\langle k \rangle  +  J_2' \ln_{\tilde{q_2}} \left( \frac{D_f}{\omega_D} N \right)$  (the intrinsic reproductive rate and cooperative term);   and $J_1'\ln_{\tilde{q_1}} \left( \frac{D_f}{\omega_D} N \right)$  (the competitive term). In this particular case,  it was assumed  
$J_2'>J_1'$ and  $q_1>q_2$. 
The lower graph presents $G(N)$ 
(which is obtained by the difference between the above two functions) 
as a function of $N$. }
\end{figure}

To show that model~(\ref{equation_macro}) can present the Allee effect, figure~(\ref{fig_allee2}) is included. 
The lower graph of this  figure  presents the form of $G(N)$  when $J_2'>J_1'$ and  $\tilde{q}_1>\tilde{q}_2$. 
In this specific case, the per- capita growth rate reaches its maximum at $N=N^*$. 
This result can be explained by analyzing  the two terms that  composed $G(N)$ in equation~(\ref{equation_macro}). The upper graphic of this figure presents the curves $\langle k \rangle  +  J_2' \ln_{\tilde{q_2}} \left( \frac{D_f}{\omega_D} N \right)$  (the intrinsic reproductive rate and cooperative term)  and $J_1'\ln_{\tilde{q_1}} \left( \frac{D_f}{\omega_D} N \right)$  (the competitive term) as a function of $N$. 
The two curves are monotonic  crescent functions of $N$, but they have  different forms. 
The per- capita growth rate $G(N)$ is the difference between these two functions. 
For small $N$, $G(N)$ is an increasing function of the population size; that is the Allee effect.

One can find the population size $N^*$ at which $G$ is at its  maximum by taking $dG(N)/dN =~0$ in~(\ref{equation_macro}), which gives
\begin{equation}\label{eq_Nestrela}
N^* = \frac{\omega_D}{D_f} \left(\frac{J_2}{J_1} \right)^{\frac{1}{\tilde{q}_1 -\tilde{q}_2 }}. 
\end{equation}
Note that if $J_2=0$, which is the MLBI model, then $N^*$ is null or indeterminate. That is, the MLBI model cannot explain Allee effect.

One can also find when  $G(N)$ becomes null, which happens at the  
the threshold value  $N=N_c$ in which the per capita growth rate became null, that is,  $G(N_c) = 0$. 
This threshold population value can be determined by solving the transcendental equation

\begin{equation}
 N_c =  \frac{\omega_D}{D_f} e_{q_1}\left[ \frac{\langle k \rangle}{J_1'} \ln_{q_2} \left( \frac{D_f}{\omega_D}  N_c\right) \right].
\end{equation}
When $N>N_c$, the population is decreasing.
Note that the maximum value of $G$ happens when the difference between the two functions plotted in the upper graph of figure~(\ref{fig_allee2}) is maximal, and the threshold value $N_c$ happens when these two functions are equal.

Model~(\ref{equation_macro}) fits very well to the muskox population  data, which are presented in figure~(\ref{fig_allee_data}). 
According to the model and result~(\ref{eq_Nestrela}), the transition from cooperation to competition for the muskox data  is $~N^*\approx246$.

\subsection{Comments}

One can interpret the Allee effect 
as cooperative and competitive interactions between the individuals of the population. 
When the population is too small, cooperation predominates, which favors the increase of the per- capita growth rate as $N$ increases. However, when the population is sufficiently large, competition predominates, which implies a decrease in the population growth rate as the population increases.

With respect to the spatial structure of the population and its representation by the model presented here, a good example is pollination in plants.  
The smaller the inter-individual distance is , the greater the efficiency of the pollination is \cite{allee_revisao, allee_ghazoul, allee_desert_mustard, allee_annual_plant}. As result, there is a cooperative effect  (or facilitation,  as argued in \cite{allee_1949})  that is strongly dependent on  the distance between the individuals.  
However,  when the inter-individual distance is small, competitive effects begin to appear in form of sunlight disputes, elimination of inhibitory toxins, or competition for soil or other resources. 
In this way, there is a  tradeoff  between the individuals to stay  close or more distant.  The Allee effect is an example of an emergent phenomenon that can emerge as a consequence of these types of  individual-individual mechanisms.

 \section{Analysis of a Special Case:  $\gamma \equiv \gamma_1 = \gamma_2$}

This section will be restricted to the special case in which the two decay exponents (for competition and cooperation) have equal values. That is,  $\gamma \equiv \gamma_1 = \gamma_2$, which is equivalent to saying that $\tilde{q} \equiv \tilde{q}_1 = \tilde{q}_2$.   
 In this particular case,  model~(\ref{equation_macro})  becomes

\begin{equation}\label{eq_J}
\frac{1}{N}  \frac{dN}{dt} =  \langle k \rangle + J'\ln_{\tilde{q}} \left(\frac{D_f}{\omega_D} N \right), 
\end{equation}
where it is assumed $J'\equiv J_2' - J_1'$. 
The parameter $J'$, which can assume both positive and negative values, determines  what type of interaction has more strength:  cooperation ($J'>0$) or competition ($J'<0$). 
When $\tilde{q}=0$, i.e. when $\gamma = D_f$,  the generalized logarithm function becomes the usual logarithm,  and then,  Eq.~(\ref{eq_J}) is the Gompertz growth model.

Using the properties of the generalized logarithm, one can show that equation~(\ref{eq_J}) can be rewritten as   

\begin{equation}\label{modelo_bertalanffy}
\frac{d}{dt} N = aN^{1+q} - bN. 
\end{equation}
with solution

\begin{equation}\label{solucao_west}
 N(t) = \left[ \frac{a}{b} + \left(N_0^{-q} - \frac{a}{b} \right) e^{bqt} \right]^{-\frac{1}{q}}.
\end{equation}
In the last two equations, the parameters $a$ and $b$ are given by

\begin{equation}
 a \equiv \frac{J'}{\tilde{q}}\left( \frac{D_f}{\omega_D} \right)^{\tilde{q}},
 \end{equation}
and 
\begin{equation}\label{definition_b}
 b \equiv \frac{J'}{\tilde{q}} - \langle k \rangle, 
\end{equation}
respectively.
Model~(\ref{modelo_bertalanffy}) is the \textit{Richards model} \cite{richards_59}, which is utilized in \cite{bertalanffy-model, savageau-1979, west-nature} to describe animal growth. Recently,  this model was studied by West et al  in the context of the growth of cities \cite{west-pnas}. Thus, the Richards model is the same model that was deduced here from the individuals' interactions.

The sign  of the argument of the exponential in~(\ref{solucao_west}) is important, as it determines the convergence (or divergence) of the population. When $b\tilde{q}<0$, the exponential term goes to zero at $t\to \infty$,  and then,  the population has a saturation. However, when $b\tilde{q}>0$ the exponential diverges. Thus,  there is a change in the behavior when $b\tilde{q}=0$, which happens when $\gamma = \gamma^*$, where

\begin{equation} \label{eq_gamma_estrela}
\gamma^* \equiv D_f \left(1- \frac{J'}{\langle k \rangle}\right)
\end{equation}
(according to the definitions in~(\ref{definition_b}) and~(\ref{definition_q})).

The term $\gamma^*$ plays an important role in the analysis of the population growth behavior,  and it will be discussed in more detail in the next section. 
The sign of the exponent  $(-1/\tilde{q})$ in~(\ref{solucao_west}) is also important in the analysis of the dynamics. 
Transition behavior happens when $1/\tilde{q} =0$, that is, when   $\gamma = D_f$ (according to~(\ref{definition_q})). 
In the next section, the analysis of solution~(\ref{solucao_west}) 
for the predomination of cooperation and competition 
 will be presented separately.  
 The convergence or divergence of the population according to the value of $\gamma$  will also be analyzed there.

\subsection{The Predominance of Cooperation}

Let's analyze the particular case in which the system described by  model~(\ref{eq_J})  presents cooperation predominance, that is, when  $J'>0$.
In this case, given that  $\langle k \rangle$ is positive, the population always grows, without saturation. However,  the way the population grows depends on the value of the exponent decay.

\begin{figure}[htbp]
\centering
\includegraphics[scale=0.33]{plano_Jpos.eps} 
\includegraphics[width=\columnwidth]{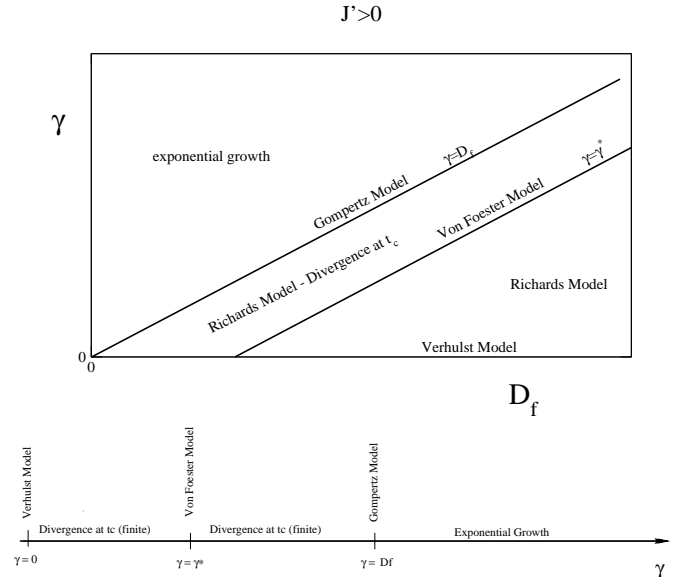} 
\caption{\label{figure_cooperation} A schema of the population growth according to the exponent decay $\gamma$  when cooperation predominates ($J'>0$). 
The upper graph is the phase diagram corresponding to  $\gamma$-\textrm{x}-$D_f$; the lower graph  presents the population growth behavior as a function of $\gamma$, where $D_f$ is fixed. 
The exponent decay represents the following models: $\gamma=0$ implies that one has the Verhulst model; $\gamma=\gamma^*$ implies one has Von Foerster model; $\gamma = D_f$ implies one has the Gompertz model. When $\gamma< D_f$, one has the Richards model, and hence, the population diverges at a finite time $t_c$ given by~(\ref{eq_tc}). When  $\gamma= D_f$ (the Gompertz model), the population diverges as  $t\to \infty$. 
lastly, when $\gamma> D_f$,  the population grows  exponentially (the Malthus model), as in Eq.~(\ref{eq_exponential}).}
\end{figure}

For instance, when $\gamma >  D_f$,  then $b<0$, $- 1/\tilde{q}>0$ and  $b\tilde{q}>0$. 
The solution for this case~(see~(\ref{solucao_west}))  can be written as

\begin{equation}\label{eq_exponential}
 N(t \gg 1) \sim \left[ e^{b\tilde{q}t} \right]^{-\frac{1}{\tilde{q}}} = e^{ \left( \langle k \rangle - \frac{J'}{\tilde{q}} \right)   t}. 
\end{equation}
That is, $\gamma > D_f$ implies exponential growth of the population,  which is the \textit{Malthus model} with growth rate $-b = \langle k \rangle - \frac{J'}{\tilde{q}}$ (see Eq~(\ref{definition_b})). 
When $\gamma = D_f$ (the Gompertz model),   the population diverges asymptotically as $N(t \gg 1) \sim e^{e^{J'\frac{\omega_D}{D_f}t}}$.
When $\gamma<D_f$,  the population diverges at a finite time $t_c$ given by

\begin{equation}\label{eq_tc}
 t_c \equiv  \frac{-1}{b\tilde{q}\ln \left( 1- \frac{b}{a} N_0^{-\tilde{q}} \right)}.
\end{equation}
These results are summarized in figure~(\ref{figure_cooperation}).

\vskip 0.5 cm 

\subsection{The Predominance of Competition}

When competition predominates, that is, when $J'<0$,  the model described here is quite similar to the MLBI model. Thus,  the analysis of the convergence or divergence of the population is  similar to the analysis discussed and presented by D'Onofrio in \cite{donofrio}. 

When $\gamma< \gamma^*$, which implies $-1/q>0$ and\\ 
$bq<0$, solution~(\ref{solucao_west}) implies that the population  converges to a finite size - which is the \textit{carrying capacity} -  and is given by

\begin{equation}\label{eq_K}
 K =  \left( \frac{a}{b} \right)^{-\frac{1}{\tilde{q}}}  =\left[ \frac{\left(\frac{D_f}{\omega_D}\right)^{\tilde{q}}}{1- \langle k \rangle \frac{\tilde{q}}{J'} }  \right]^{-\frac{1}{\tilde{q}}}. 
\end{equation}

\begin{figure}[htbp]
\centering
\includegraphics[scale=0.33]{plano_Jnega.eps} 
\includegraphics[width=\columnwidth]{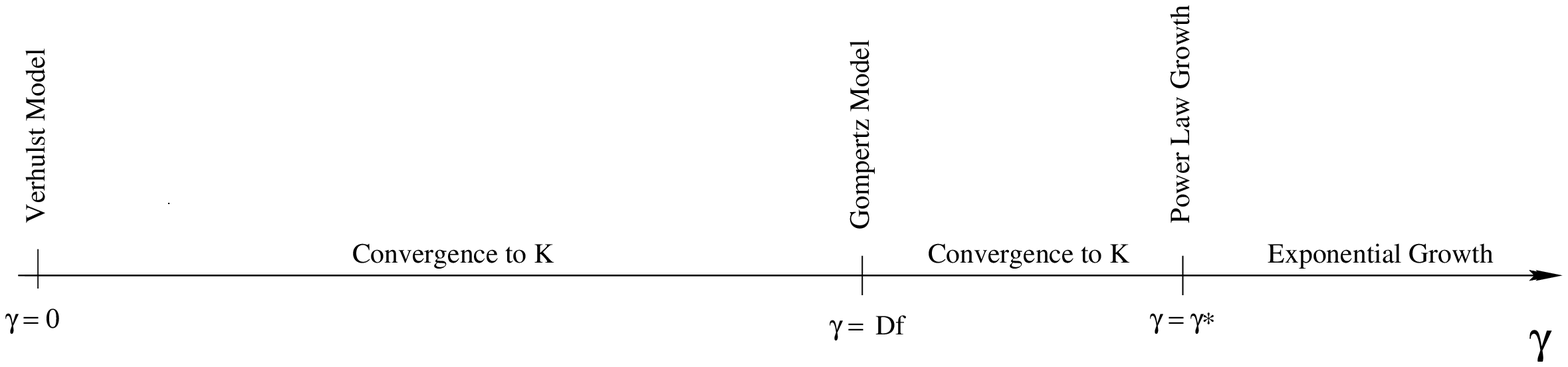} 
\caption{ \label{figure_competition} A schema of the population growth according to the exponent decay $\gamma$,  when  competition predominates ($J'<0$). whereas the upper graph is the phase diagram $\gamma$-\textrm{x}-$D_f$, the lower graph presents the population growth behavior as a function of $\gamma$, where $D_f$ is fixed. 
The exponent decay represents the following models: $\gamma=0$ implies one has the Verhulst model; $\gamma=\gamma^*$ implies power law growth  (see eq.~(\ref{eq_power})); $\gamma = D_f$ implies one has the Gompertz model. When $\gamma< \gamma^*$, the population converges to a finite size (the carrying capacity $K$), which is given by~(\ref{eq_K}). 
When $\gamma>\gamma^*$ the population presents exponential growth, which  conforms to eq.~(\ref{eq_exponential_compe}), and  the population goes to extinction when  $|J'/\tilde{q}| >\langle k \rangle$; it diverges otherwise.}
\end{figure}

When $\gamma>\gamma^*$ (which implies $\tilde{q}<0$), at $t\gg1$ 
solution~(\ref{solucao_west}) becomes    

\begin{equation}\label{eq_exponential_compe}
 N(t) \sim e^{\left(\langle k \rangle - \frac{J'}{\tilde{q}} \right)t} = e^{\left(\langle k \rangle - |\frac{J'}{\tilde{q}}|\right)t},
\end{equation}
i.e., one has exponential growth (the Malthus model).  When $ |J'/\tilde{q}| <  \langle k \rangle$ the population diverges exponentially. As the population grows indefinitely even due to  predominance of competition, one can call this situation  \textit{weak competition}. However,  when  $|J'/\tilde{q}| >  \langle k \rangle$, the population decays exponentially, and it goes to extinction for $t\to \infty$. Thus, one can call this situation \textit{strong competition}. 
These results are summarized in figure~(\ref{figure_competition}). 
The case $\gamma=\gamma^*$ will be analyzed in the next section in the  general context of the parameter $J'$ (which may be positive or negative).

\vskip 0.5 cm 

\subsection{Comments about $\gamma = \gamma*$ and the Human Population Growth}

The particular case that $\gamma = \gamma^*$ must receive more attention. 
In this case,  the parameter $b$ becomes null  (according to   Eq.~(\ref{eq_gamma_estrela}) and~(\ref{definition_b})), and then,  the model~(\ref{solucao_west})  becomes

\begin{equation}\label{modelo_vonfoester}
 \frac{d}{dt}N = aN^{1+\tilde{q}}.
 \end{equation}
This is the von Foerster growth model, which was studied both in \cite{von-foester} and more recently in \cite{polones} to describe human population growth. 
The solution of model~(\ref{modelo_vonfoester}) is

\begin{equation}\label{solution_vonfoester}
N(t) = N_0 e_{\tilde{q}}\left(\frac{a}{N_0^{\tilde{q}}}  t\right),
\end{equation}
which was presented in \cite{fabiano-physicaA, fabiano-pre},   
where $e_q(x)$ is the generalized exponential function (see appendix~\ref{appendix_lnq}). 
Note that result~(\ref{solution_vonfoester}) is exactly the model proposed in  \cite{polones_q}.  In this reference, the model was introduced by a modification of the exponential term of the Malthus model solution without any justification. However, with the formulation of the microscopic model proposed here,  all of the involved quantities have a physical interpretation, and the growth behavior described by~(\ref{solution_vonfoester}) is a consequence of the of interactions of the individuals.

Given that  $\omega_D$, $D_f$,  and $\langle k \rangle$ are positive parameters, the manner in which the population grows for large $t$ is totally dependent on $J'$. 
For instance, when $J'=0$,  the population grows exponentially  because  the competitive and cooperative strength completely cancel each other out,  and then,  the population grows without individual interactions.

\begin{figure}[htbp]
\centering
\includegraphics[width=\columnwidth]{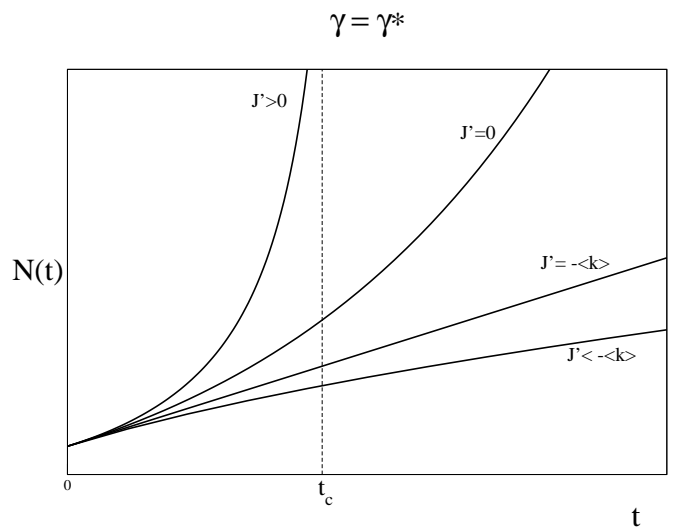} 
\caption{ \label{figure_N} 
The time evolution of the population according to the strength of the interaction $J'$ for $\gamma = \gamma^*$. 
When $J'>0$, that is, when one has cooperation, the population diverges at a finite time $t_c$ (in the Von Foerster model). When $J'=0$,  the population experiences exponential growth. In this case,  the interaction effect is fully nullified and the growth rate is given only by the intrinsic growth rate $\langle k \rangle$. When $J'<0$, that is, when one has competition, the population diverges only as  $t\to \infty$. The special cases are: $J'= -\langle k \rangle$,  linear growth; and  $J'< -\langle k \rangle$, logarithmic growth.
$N(t)$ is a convex function when $J'> - \langle k \rangle$, and it is a concave function otherwise. }
\end{figure}

When $J'<0$, solution~(\ref{solution_vonfoester}) behaves asymptomatically as a power law:  

\begin{equation}\label{eq_power}
 N(t \gg 1) \sim t^{- \frac{\langle k \rangle}{J'}}.
\end{equation}
In this way, if $J'<0$,  then  the population diverges only when $t \to \infty$. 
The concavity of $N(t)$ is also determinede by $J'$: if   
$J'> - \langle k \rangle$, then $N(t)$ is a convex function, and it is  concave otherwise. 
Whereas $J'<0$  the population diverges only when $t \to \infty$, for $J'>0$ the population diverges at a finite time $t_c$, which is given by

\begin{equation}
 t_c = \frac{1}{J'} \left(\frac{N_0 D_f}{\omega_D}\right)^{- \frac{J'}{\langle k \rangle}}.
 \end{equation}
Figure~(\ref{figure_N}) summarizes these conclusions.


\begin{figure}[htbp]
\centering
\includegraphics[width=\columnwidth]{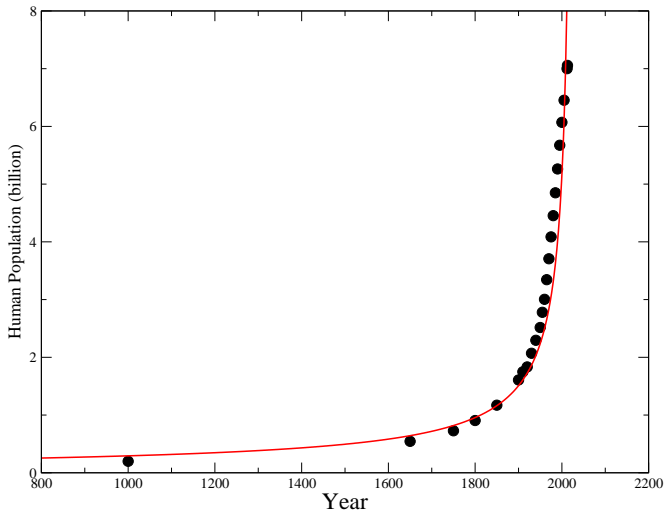} 
\caption{\label{human_fig}
The human population as a function of time since the Middle Ages. The data were obtained from \cite{polones} and from the
U.S. Census Bureau \cite{census}. The curve is a plot of the equation 
$N(t) =  65,6(2026 -t)^{-0,78}$  (from Eq.~(\ref{eq_power_law_growth})), whose parameters values were  obtained via a data fit. 
}
\end{figure}

When $J'>0$, it is interesting to write  solution~(\ref{modelo_vonfoester}) in terms of the critical time $t_c$ in which the population diverges.Thus, solution~(\ref{solution_vonfoester})  behaves as

\begin{equation}\label{eq_power_law_growth}
 N(t) \sim (t_c-t)^{- \frac{\langle k \rangle}{J'}}
\end{equation}
when cooperation predominates. 
An interesting application of this result is in human population growth, as represented in figure~(\ref{human_fig}). Note that  Eq.~(\ref{eq_power_law_growth}) applies very well to human population growth, as the data from $1000$ A.D.  until $2013$ conform with what  was presented in \cite{polones, von-foester}. However, with the presentation of the microscopic point of view of the interactions between the individuals, one can argue that the ``divergent behavior'' of the human population can be seen as a result of cooperative effects, as the parameter $J'$ must be positive (i.e. cooperation predominates) to fit the data.

%
%

\section{Conclusion}

In the present work, one has proposed a growth model based on the microscopic level of the interaction of the individuals that  constitute the population. It was shown that the model reached  several well known growth models presented in literature as special cases. For instance, one has obtained the Malthus, Verhulst, Gompertz, Richards, Bertalanffy, power law, and Von Foerster growth models. 
The present model explains some macroscopic behaviors using a 
non-~phenomenological approach. 
Moreover,  it uses parameters that have physical meanings and that  can 
be measured in real systems. 
In addiction,  the model showed more flexibility than the original version (i.e., the MLBI model). 
For instance, the extended model presented here permits us to explain  the Allee effect as an emergent behavior from the individual-individual interactions, which is contrast to the common phenomenological explanation presented in the literature. 
It is important to stress that the MLBI model, which considers only competitive  interactions, can not explain this effect.

It was observed that the relation between the decay exponent ($\gamma$), the fractal dimension ($D_f$) of the population, and the interaction strength ($J$) determine the behavior of the population growth. 
For instance, one has presented a phase diagram in which one related  diverse types of growth as consequences of the distance dependent interactions (by  the exponent decay $\gamma$) and the fractal dimension of the population.
Moreover, one has shown how the strength of the interaction gives both the concavity of the growth (as a function of time) and the saturation or divergence of the population.

In conclusion, the model proposed here incorporates many types of macroscopic ecological patterns by focusing on the balance of cooperative and competitive interactions at the individual level.
In this way, the model presents a  new direction in the search for  universal patterns,   which could shed more light on population growth behavior.

\section*{Acknowledgements}
I would like to acknowledge the useful and stimulating  discussions with Alexandre Souto Martinez and Brenno Troca Cabella.  

\newpage

\appendix

\section{The Generalized Logarithm and Exponential Function} \label{appendix_lnq}

In this appendix, one presents the generalizations of the 
logarithmic and exponential functions and  some of their properties.
The introduction of the functions is shown to be very useful for dealing with the mathematical representation of the population growth model that is presented in this work.  

The $\tilde{q}$-\textit{logarithm function} is defined as
\begin{equation}
\label{definition-lnq}
\ln_{\tilde{q}}(x) = \lim_{\tilde{q}' \rightarrow \tilde{q}}\frac{x^{\tilde{q}'} -1 }{\tilde{q}'} = \int_1^x \frac{dt}{t^{1-\tilde{q}}}  \; ,
\end{equation}
which is the area of the crooked hyperbole, and is controlled by $\tilde{q}$.
This equation is a generalization of the natural logarithm function, which is  reproduced when $\tilde{q} = 0$. 
This function was introduced in the context of nonextensive statistical mechanics \cite{tsallis_1988, tsallis_qm} and was studied recently in \cite{arruda_2008, martinez:2008b, Martinez:2009p1410}. 
Some of the properties of this function are as follows: for $\tilde{q} < 0$, $\ln_{\tilde{q}}(\infty)=-1/\tilde{q}$; for $\tilde{q} > 0$, $\ln_{\tilde{q}}(0)=-1/\tilde{q}$; for all $\tilde{q}$, $\ln_{\tilde{q}}(1)=0$; $\ln_{\tilde{q}}(x^{-1}) = - \ln_{-\tilde{q}}(x)$; $d \ln_{\tilde{q}}(x)/dx = x^{\tilde{q}-1}$. Moreover, the $\tilde{q}$-logarithm is a function:
convex for $\tilde{q}>1$; linear for $\tilde{q}=1$; and concave for $\tilde{q}<1$.

The inverse of the $\tilde{q}$-\textit{logarithm function}  is the $\tilde{q}$-\textit{exponential function}, which is by
\begin{equation}
e_{\tilde{q}}(x) =   
\left\{ \begin{array}{ll}
\lim_{\tilde{q}^{'} \to \tilde{q}}   (1+ \tilde{q}^{'} x)^{ \frac{1}{ \tilde{q}^{'}} } & ,\textrm{ if $\tilde{q}x > -1$} \\
0 & ,  \textrm{ otherwise} 
\end{array} \right. \; .
\label{def-eq}
\end{equation}
Some properties of this function are as follows:  $e_{\tilde{q}}(0)=1$, for all $\tilde{q}$;  $ \left[ e_{\tilde{q}}(x)  \right]^a = e_{\tilde{q}/a}(ax)$, where $a$ is a constant; and  for $a=-1$, one has $1/e_{\tilde{q}}(x) = e_{-\tilde{q}}(-x)$.
Moreover, the $\tilde{q}$-exponential is a function:
convex for $\tilde{q}<1$; linear for $\tilde{q}=1$; and concave for $\tilde{q}>1$.

\section{A Detailed Calculus of $I_i^{(l)}$}\label{appendix_I}

In this appendix one presents a detailed calculus for the intensity of the interaction felt by a single individual $i$ from the other individuals of the population, which is represented by $I_i^{(l)}$  (see section~(\ref{section_model}). One follows Mombach et al \cite{mombach} to show that this intensity is independent of the individual. That is, it is the same for all individuals of the population and depends only on the size of the population. More specifically, one shows that $I_i^{(l)} = I^{(l)}(N)$ regardless of $i$.

First, it was presented in section~(\ref{section_model}) that 

\begin{equation}\label{Ii}
 I_i^{(l)} =  \sum_{j\ne i } \frac{1 }{|\mathbf{r}_i-\mathbf{r}_j |^{-\gamma_l}}    = \sum_{j=1}^N \frac{\left(1-\delta_{ij}\right)}{|\mathbf{r}_i-\mathbf{r}_j |^{-\gamma_l}},  
\end{equation}
where $\delta_{ij}$, which is the Kronecker's delta, was introduced to avoid the restriction in the sum. 
Introducing the property 

\begin{equation}
 f(\mathbf{r}_0) = \int_{V_D} d^D\mathbf{r} \delta(\mathbf{r} - \mathbf{r}_0) f(\mathbf{r}),
\end{equation}
where $\delta(\cdots)$ is the Dirac's delta, the expression~(\ref{Ii}) becames

\begin{equation}\label{delta-intro}
I_i^{(l)} = \sum_{j=1}^N (1-\delta_{ij}) \int_{V_D} d^D\mathbf{r} \delta\Big(  \mathbf{r} - (\mathbf{r}_j - \mathbf{r}_i)\Big) |\mathbf{r}|^{- \gamma_l}.
\end{equation}
In the last two expressions, was introduced:   $D (=1,2, 3)$, which is   the Euclidean dimension in which the population is embedded, and  $V_D$, which is the total (hipper)volume (in $D$ dimensions) that  contains the population. 
The form represented in~(\ref{delta-intro}) was obtained by the  variable substitution $\mathbf{r}_j - \mathbf{r}_i$ by $\mathbf{r}$, using Dirac's delta.

Some algebraic manipulation and the introduction of $r\equiv|\mathbf{r}|$, allows to write  

\begin{equation}\label{value_I}
 I_i^{(l)} =\int_{V_D}\frac{d^D \mathbf{r}  }{\mathbf{r}^{\gamma_l}} \sum_{j \ne i} \delta\Big(  \mathbf{r} - (\mathbf{r}_j - \mathbf{r}_j)\Big).
\end{equation}

Note that $dN(\mathbf{r})\equiv d^D  \mathbf{r} \sum_{j \ne i} \delta\Big(  \mathbf{r} - (\mathbf{r}_j - \mathbf{r}_j)\Big)$ is the  number of individuals which is at the element of (hipper)volume $d^D\mathbf{r}$ at the distance $\mathbf{r}$ from the individual $i$, localized at $\mathbf{r}_i$.  In this way, the  density of individuals  at $\mathbf{r}_i + \mathbf{r}$ (neighbors of $i$), that is $\rho(\mathbf{r}_i~+~\mathbf{r}) =dN(\mathbf{r})/d^D \mathbf{r}$, can be written as 

\begin{equation}\label{eq_rho}
 \rho(\mathbf{r}_i + \mathbf{r}) =  \sum_{j \ne i} \delta\Big(  \mathbf{r} - (\mathbf{r}_j -\mathbf{r}_j)\Big).
 \end{equation}

The density of individuals can also be thought of in terms of the scale of the system (in conformity with \cite{Falconer}). 
The volume of the system grows in the form  $V_D \sim L^D$, where 
 $L$ is the typical size of the system. However, the number of individuals grows as the form $N \sim L^{D_f}$, where $D_f$ is the fractal dimension formed by the spatial structure of the population.
By considering $r$, which is the absolute distance from $i$,  as a typical distance of the system, one can say that the density of individuals ($V_D/N$) has the form 

\begin{equation}\label{rho_value2}
\rho(\mathbf{r}_i + \mathbf{r}) \equiv \rho(r) = a \frac{r^{D_f}}{r^D},
\end{equation}
where $a$ is a constant. 

Using results~(\ref{rho_value2}) and~(\ref{eq_rho}) in~(\ref{value_I}), one obtains

\begin{equation}
 I_i^{(l)} = a \int_{V_D}d^D \mathbf{r} r^{D_f -D -\gamma_l}. 
\end{equation}
Note that the integration argument does not depend on the angular coordinates. Thus, one can write  
$d^D \mathbf{r}=~r^{D-1}dr d\Omega_D$, where $d\Omega_D$
is the solid angle, which implies 

\begin{equation}
 I_i^{(l)} = a \int d\Omega_D \int_{r_0}^{R_{max}} dr r^{D_f -1-\gamma_l} 
\end{equation}
Note that the only term that depends on the Euclidean dimension is the solid angle, and the integral $\int d\omega_D$ assumes the following values according to these  tree possibilities: $D=1$,  $\int d\omega_1 = 1$;  $D=2$,  $\int d\Omega_2 = 2\pi$;  $D=3$,  $\int d\Omega_3 = 4\pi$. 
By introducing the constant $\omega_D = a \int d\Omega_D$, which depends only on $D$,  one obtains 
\begin{equation}\label{equation_I}
I^{(l)} \equiv I_i^{(l)} = \omega_D \left( \frac{R_{max}^{D_f - \gamma_l} -1 }{D_f - \gamma_l}\right)
\end{equation}
Thus, $I_i^{(l)}$  does not depend on the label $i$ anymore. As a result, one can say that $I_i^{(l)} = I^{(l)}$ regardless of $i$. 

Furthemore, one can introduce the total number of individuals in the relation above by the following thinking. 
The total number of individuals in the population can be determined by the integral 

\begin{equation}
 N = \int dN(r) =\int_{V_D}d^D\mathbf{r} \rho(r). 
 \end{equation}
Using equation~(\ref{rho_value2}) and integrating the solid angle, one obtains   

\begin{eqnarray}
N &=& \omega_D \int_{0}^{R_{max}} r^{D_f-1}dr  \label{eq1} \\
&= &\omega_D \int_{0}^{r_0} r^{D_f-1} + \omega_D \int_{r_0}^{R_{max}} r^{D_f-1}  \label{eq2}\\
&= & \omega_D \frac{r_0^{D_f}}{D_f} +  \omega_D \frac{R_{max}^{D_f}}{D_f}
 - \omega_D \frac{r_0^{D_f}}{D_f} \label{eq3}.
\end{eqnarray}
Note that the first term on the right in~(\ref{eq2}) and~(\ref{eq3}) can be zero (indicating the absence of individuals) or 1 (indicating the presence of a single individual). These values are possible because the ratio of the individual is $r_0$,  and hence, there can be at most one individual inside the region that consists of the length between $0$ and $r_0$. Thus, for $r_0=1$, $\omega_d/D_f \sim  1$.
$R_{max}$ can  be obtained from~(\ref{eq3}), which is a function of $N$ according to  

\begin{equation}
 R_{max}  = \left( \frac{D_f}{\omega_D} N  \right)^{\frac{1}{D_f}}. 
\end{equation}
Returning to relation~(\ref{equation_I}) one finds  

\begin{equation}
 I = I(N) = \frac{\omega_D}{D_f(1- \frac{\gamma}{D_f})}\left[ \left(\frac{D_f}{\omega_D}N \right)^{1- \frac{\gamma}{D_f}} -1 \right].
\end{equation}
By introducing $\tilde{q}=1- \gamma/D_f$ and the properties of the generalized logarithm (Appendix~\ref{appendix_lnq}) one obtains

\begin{equation}
I = I(N| D, \tilde{q}) = \frac{\omega_D}{D_f} \ln_{\tilde{q}} \left(\frac{D_f}{\omega_D}N \right). 
\end{equation}


\end{document}